# Self-Assembled InAs Quantum Dots on Patterned InP Substrates

J. Lefebvre, P.J. Poole, J. Fraser, G.C. Aers, D. Chithrani, and R.L. Williams

Institute for Microstructural Sciences, National Research Council Canada, Ottawa (Ontario), K1A 0R6, Canada

## ABSTRACT

The size distribution of self-assembled InAs quantum dots grown on (001) InP under the Stranski-Krastanow growth mode is controlled using selective area/chemical beam epitaxy, which allows the formation of quantum dots at specific locations. As the dimensions of the patterned areas are decreased from 1000 nm down to 50 nm or less, scanning electron microscopy reveals a gradual increase in the spatial correlation between quantum dots, which leads to the formation of ordered arrays for dimensions below 200 nm.

## INTRODUCTION

Fabrication methods relying on a self-assembly process already show great promise in fulfilling many requirements of the future nano-electronics and the nano-optics industries. In many cases, these self-assembly processes are inspired by biological systems and rely upon the chemical affinity of particular groups. In other cases, physical properties favor a specific size, shape and configuration for clusters of atoms, molecules etc. Chemical self-assembly techniques are found to produce virtually identical replicas, while physical self-assembly most often presents a significant spread around a required characteristic. This latter behavior is observed following the Stranski-Krastanow growth of self-assembled semiconductor quantum dots (SADs) on crystalline substrates. The large spread in the size distribution of SADs grown under the Stranski-Krastanow growth mode is an important drawback for many future applications such as ultra low threshold lasers and qbit architectures for solid state, scalable quantum computers.

Optical and electrical measurements performed on individual SADs reveal the zero-dimensional character that makes quantum dots appealing for both exploratory physics and future applications [1]. However, measurements performed on a macroscopic scale show that many of these features are washed out because such measurements provide a spatial average over quantum dots of many different sizes and electronic configurations. To fully realize the device potential of self-assembled quantum dots, the size distribution and the resulting inhomogeneity in electronic characteristics must be reduced significantly. In this paper, we address this issue experimentally, and show that SADs can be grown on nano-structured templates to produce ordered arrays as the template width is reduced below 200 nm.

## EXPERIMENTAL DETAILS

Narrower size distributions have been obtained previously by growing SADs at specific locations using selective area epitaxy. Here, we choose a similar method and pattern the exactly oriented InP (001) substrates ex-situ by etching (reactive ion etching) openings in a $SiO_2$ layer deposited on the

*Figure 1*. Scanning electron microscopy micrographs of InP mesas grown using selective area/chemical beam epitaxy. The InP stripes oriented along [100], [110], and [1$\bar{1}$0] directions are grown inside openings etched in a SiO$_2$ mask (grainy surface surrounding the stripes).

InP substrate. The openings, defined by electron beam lithography, consist of 200 to 1000 nm wide lines running along the [100], [110], and [1$\bar{1}$0] directions. Prior to growth, a slow InP wet etch (H$_3$PO$_4$:H$_2$O$_2$:H$_2$0 1:1:10, 3 nm/min) is used to remove possible surface contaminants. Chemical beam epitaxy (CBE) with tri-methyl-indium, arsine, and phosphine is used to grow InP templates and InAs SADs. The growth first proceeds by depositing 70-150 nm of InP at 500 °C at 0.5 µm/hour growth rate. During InP template growth and subsequent dot deposition, growth only occurs inside the patterned openings, and low index facets are generated spontaneously to produce stripes with a trapezoidal cross-section and a (001) top surface. The InAs SADs are grown on the (001) surface by depositing between 1.5 and 2.1 monolayers (ML) of InAs. The structure is left uncapped for scanning electron microscopy (SEM) and atomic force microscopy (AFM) studies. In order to prevent the formation of large InAs clusters, the InAs deposition is immediately followed by rapid cooling under a low arsenic overpressure (four times lower than for the growth of capped quantum dots). This procedure differs from that used for capped samples, where a 15 to 30 sec growth interruption is used following InAs deposition, followed by growth of an InP cap. More details on the growth of InP and InAs SADs on unpatterned substrates can be found elsewhere [2].

**DISCUSSION**

Figure 1 shows stripes running along the [100], [110], and [1$\bar{1}$0] directions. For the growth conditions used here, the [110] and [100] directions produce well developed, virtually defect-free side facets. In contrast, stripes oriented along the [1$\bar{1}$0] direction produce side facets with a substantial number of defects. The [100] direction is characterized by the formation of (011) facets at a 45° angle with respect to the top (001) facet. The appearance of low index side facets along the mesa edges may be viewed as the result of a varying growth rate for the various crystallographic planes. More specifically, the growth rate is lower on low surface energy planes, producing a large population of surface adatoms available for surface diffusion to adjacent, higher growth rate facets. In such a picture, In atoms would migrate from the (011) side facets towards the (001) top facet, denuding the side facet of InAs source material for quantum dot formation, as described later. The diffusion of source material away from the low growth facets is used here to reduce the lateral dimensions of the mesa, and to produce nano-scale templates for the growth of SADs. The templates are expected to be virtually free from process induced defects since they are formed entirely during the growth process. A perfect ridge structure can even be obtained from a mask with imperfect edges. As can be seen in figure 3, the self-smoothing of side facets give rise to a well defined (001) plane with a uniform width that extends over tens of micrometers, even though the mask roughness is of the order of a few tens of nanometers.

*Figure 2*. The stripe top width depends critically on the patterned width. The continuous line is the result of equation (1) with 45 nm of incorporated InP, which is half the amount deposited on a planar substrate.

*Figure 3: scanning electron microscopy images of InAs SADs grown on mesa tops of different width. The reduction in mesa top width results in an increased spatial correlation between SADs and to the formation of linear chains of SADs. The stripes are along the [010] direction.*

The width of the mesa top depends critically on the original width of the patterned opening and on the amount of deposited material. Figure 2 shows the decrease of the top width (T) as the patterned width (W) is decreased for a constant amount of deposited material (100 nm), as determined from the growth rate on planar substrates. The result is obtained for a single growth run with a sample patterned with stripes of different width. The reduction in top width can be predicted using simple geometrical constructions. If we consider that (011) side facets form readily, and that the deposited InP (cross-section = Wh) is incorporated into the mesa structure, we find that "T", "W", and the thickness of material incorporated "h", are related simply by,

$$T = (W^2 - 4Wh)^{1/2} \quad (1).$$

The data are fitted well using h=45 nm, a value corresponding to about half the deposited amount. This equation predicts that the top (001) facet should vanish for a 180 nm wide stripe. However, although equation (1) agrees well with the experimental data, we find that for our growth conditions, fully completed ridges never form. Instead, an approximately 20 nm wide top remains even after a prolonged growth. Although this characteristic of the growth is extremely interesting, its discussion is deferred to a later publication, as it requires a more extensive study.

Figure 3 shows a series of stripes of different widths, following the deposition of 1.5 ML of InAs. SADs are observed only on the top (001) surface of the templates. The amount of deposited InAs was chosen to be below the critical thickness of 2.5 ML obtained for SAD formation on unpatterned substrates [2]. The formation of SADs is understood as a consequence of significant amounts of In diffusing away from the side facets in favor of the top (001) mesa surface, thus increasing the amount of InAs above the critical thickness. SADs are clearly seen on each of the ridges shown in figure 3. The surprisingly high contrast obtained from the SEM is due to topography, since atomic force microscopy (AFM) reveals that uncapped SADs are between 8 and 15 nm tall (figure 4), depending on growth conditions. These heights are extremely large in comparison with the estimate of 1.5-4 nm obtained from transmission electron microscopy (TEM) and photoluminescence experiments performed on capped samples [2]. In addition, comparison between SEM and TEM shows that the SADs have a slight tendency to form bigger clusters when left uncapped.

*Figure 4. AFM micrograph of a linear array of InAs SADs on InP. Depending on the growth conditions, SADs are 8 to 15 nm in height.*

In figure 3, a set of three samples with different stripe widths clearly shows the onset of spatial ordering to produce single or double SAD chains. The growth of SADs on narrow stripes has already proven successful in producing ordered arrays of SADs [3-5]. One-dimensional ordering was attributed to a dimensional crossover in surface diffusion as the stripes are made narrower than the lateral diffusion length. For InAs on InP, micrometer-long diffusion lengths have been measured, confirming that one-dimensional ordering should occur in this system for the stripe widths used here [6,7]. Although spatial ordering has been achieved in the work described here, further optimization of the growth conditions will be necessary to reduce the number of vacancies along the chain. Such optimization should also allow the formation of ordered quantum dots with increased co-ordination number and ultimately the production of ordered 2D lattices.

As described above, the formation of SADs on the mesa top relies on the diffusion of adatoms away from the side facets and onto surrounding surfaces. For wide stripes, this amount of material is not significant in comparison to that directly deposited. If a sub-critical layer of InAs is deposited on a planar substrate, no SADs are expected on the top of wide mesas. In figure 5, the SAD surface density is shown as a function of mesa top width. A sudden, six-fold increase is observed when the top width is decreased below 300 nm. This onset can be translated into an effective amount of InAs on the mesa top, calculated simply as the amount of InAs deposited, enhanced by the mesa base to top width ratio (W/T). We find the onset at approximately 2.1 ML, a value comparable to the 2.5 ML critical thickness measured on unpatterned substrates [2].

The SAD spatial ordering induced by changes in surface diffusion characteristics is expected to affect SAD size distribution. In an ideal case, where the distance between SADs is perfectly regular, the amount of material supplied to each SAD during the growth is necessarily uniform, making each SAD identical. For SADs grown on planar substrates, long range spatial ordering is difficult to achieve due to the presence of seeding points such as atomic layer steps or surface defects. Moreover, spatial ordering is driven by the formation of an InAs depletion area surrounding each SAD. In two dimensions, the depletion area may not be well defined since material diffuses in all directions. As the SADs are forced to locate on small mesa tops, the depletion area becomes better defined and the density of seeding points decreases. Consequently, a more regular spatial ordering is favored in one dimension, which in turn produces a more uniform SAD size distribution. Although the results are preliminary, we have some indications of an improved size distribution for our ordered linear arrays of SADs. Photoluminescence measurements will give more quantitative results of the effect of spatial ordering on the size distribution of SADs.

*Figure 5. Surface density of SADs as a function of mesa top width and effective amount of deposited InAs.*

**CONCLUSION**

In this work, we have shown that defect free templates fabricated using selective area/chemical beam epitaxy can be used to control the location of InAs SADs on InP substrates. Using this method, linear chains of ordered SADs have been obtained. Spatial selectivity is also achieved as the SAD surface density is controllably modulated by two orders of magnitude on the same sample.

Finally, preliminary results indicate that SADs in ordered arrays have a more uniform size distribution.

**ACKNOWLEDGEMENTS**

We would like to thank J. Lapointe and M. Buchanan for their help in the cleanroom, and S. Moisa for the AFM work. D.C. is supported by an NSERC fellowship.